\documentstyle[prl,multicol,aps,psfig]{revtex}
\def \beq{\begin{equation}}
\def \eeq{\end{equation}}
\def \beqa{\begin{eqnarray}}
\def \eeqa{\end{eqnarray}}
\begin{document}
\draft
\title{Quark number susceptibilities, strangeness and dynamical confinement}
\author{Rajiv V.\ Gavai \cite{ervg}}
\address{Department of Theoretical Physics, Tata Institute of Fundamental
         Research,\\ Homi Bhabha Road, Mumbai 400005, India.}
\author{Sourendu Gupta \cite{esg}}
\address{Physics Department, Brookhaven National Laboratory, P.O.Box 5000,\\
         Upton, New York 11973-5000, USA.}
\maketitle
\begin{abstract}
We report first results on the strange quark number susceptibility,
$\chi_s$, over a large range of temperatures, mainly in the plasma
phase of QCD.  $\chi_s$ jumps across the phase transition temperature,
$T_c$, and grows rapidly with temperature above but close to $T_c$.
For all quark masses and susceptibilities in the entire temperature range
studied, we found
significant departures from ideal-gas values.  We also observed a strong
correlation between these quantities and the susceptibility in the
scalar/pseudo-scalar channel, supporting ideas of ``dynamical
confinement'' in the high temperature phase of the QCD plasma.
\end{abstract}
\pacs{11.15.Ha, 12.38.Mh\hfill TIFR/TH/01-12, BNL-NT-01/5, hep-lat/0103013}

\begin{multicols}{2}
As seen recently in the Quark Matter 2001 conference \cite{qm01}, the
first runs of the RHIC have provided many exciting results. With the
anticipated much longer runtime, at a variety of colliding energies and
nuclear species, the existence and properties of the high temperature
phase of quantum chromodynamics (QCD) will be probed and tested further.
This is an immediate reason to improve our theoretical knowledge of such
matter.

Fully non-perturbative computations of quark number susceptibilities
are important for three reasons. Firstly, there have been attempts to
link them directly to experimental measurements of event-to-event
fluctuations in particle production \cite{koch}. Secondly, experimental
observations of a relative enhancement of strange quarks have been
attributed to the formation of a QCD plasma \cite{pbm}; a hypothesis
which can be quantitatively tested against the computation of the
strange quark susceptibility. Finally, earlier results
\cite{gott,gavai,bern,milc} showed a strong jump in the susceptibility
across the phase transition, but indicated a statistically significant
20\% departure from weak-coupling behaviour; this is the only (lattice)
evidence that physics at finite chemical potential is not weak-coupling
physics. In this paper we present the first extensive systematic study
of such susceptibilities and relate them to other widely studied
variables.

The partition function of QCD with three quark species can be written on
a discrete space-time lattice as
\beqa
\nonumber
   Z(T,\mu_u,\mu_d,\mu_s) &=& \int_{U,\psi,\bar\psi} \exp[-S(T)]\times\\
       &&\quad\prod_{f=u,d,s}\det M(T,m_f,\mu_f),
\label{part}\eeqa
where the temperature $T$ enters through the size of the Euclidean time
direction, $S(T)$ is the gluonic part of the action, the determinants
of the Dirac matrices, $M$, contain as parameters the quark masses, $m$,
and the chemical potentials $\mu$.  The chemical potentials for
specific flavours, $\mu_f$ ($f=u,d,s$), can be expressed as linear
combinations of the chemical potentials, $\mu_\alpha$ ($\alpha$=0,3,8),
corresponding to the diagonal generators of flavour $SU(3)$. We work
with the choice $\mu_0=\mu_u+\mu_d+\mu_s$, $\mu_3=\mu_u-\mu_d$ and
$\mu_8=\mu_u+\mu_d-2\mu_s$, because $\mu_0$ is then the usual baryon
chemical potential.

We follow convention in defining quark numbers
\beq
   n_i(T,\mu_u,\mu_d,\mu_s) = {T \over V} \frac{\partial\log Z}{\partial\mu_i},
\label{numb}\eeq
and number susceptibilities
\beq
   \chi_{ij}(T,\mu_u,\mu_d,\mu_s) =
    {T \over V}  \frac{\partial^2\log Z}{\partial\mu_i\partial\mu_j}.
\label{susc}\eeq
Here $V$ denotes the spatial volume, and $i$ and $j$ are either of the
index sets $f$ or $\alpha$. Transformations between these bases are
straightforward.  We lighten our notation by writing the diagonal
susceptibilities $\chi_{ii}$ as $\chi_i$.  We determine the
susceptibilities at zero chemical potentials, $\mu_f=0$.  In this
limit, of course, $n_i(T)=0$ for all $i$. It is also easy to prove that
$\chi_{03}=0$ if $m_u=m_d$. Since all current lattice computations are
made in this approximation, the off-diagonal susceptibility cannot be
measured at present.  Also, in this limit, $\chi_{us}=\chi_{ds}$, where
the strange quark mass, $m_s\ne m_{u,d}$.

Previous works have reported the values of the flavour $SU(2)$ singlet,
$\chi_0$, and triplet, $\chi_3$, susceptibilities. Measurements of
$\chi_{0,3}$ in 2-flavour dynamical QCD showed a statistically
significant 20\% departure from free field theory (FFT) for $m/T=0.1$
and 0.15 \cite{gott,bern}.  Later computations with $m/T=0.05$ reported
$\chi_{0,3}$ compatible with FFT but with errors of about 20\%
\cite{milc}. Quenched computations deviated from the first set of
dynamical QCD results by 5--10\% \cite{gavai}. All this work
covered a range of temperatures up to $1.5T_c$. However, the quark mass
varied with temperature, since the quantity that was fixed was $m/T$.

In order to facilitate a comparison of our results with these, we
report the same susceptibilities. In the remainder of this paper,
$\chi_0$ and $\chi_3$ refer to these 2-flavour quantities. We also
report the first determination of the strange quark susceptibility,
$\chi_s$.  This work improves on previous studies in four other ways---
by covering a larger range of temperatures, by using a series of
different quark masses at fixed $m/T_c$, by investigating finite
spatial volume effects systematically, and by analysing the effect of
increasing statistics in the stochastic determination of Fermion
operators on the lattice. The study of volume dependence gives control
over extrapolation to the thermodynamic limit.  Our study of many
different quark masses gives the strange quark susceptibility. The
systematic study of the stochastic method yields a vastly improved
determination of the flavour-singlet susceptibility, $\chi_0$.

These computations have been made on lattices with lattice spacing
$a=1/4T$ in the quenched approximation. Fermion loops are neglected in
this approximation, making it substantially easier to handle
numerically.  Apart from an overall normalisation of the temperature
scale, this approximation is known to reproduce all the qualitative
features of the full QCD simulations. Furthermore numerical agreement
between the quenched and 2-flavour dynamical QCD results for
$\chi_{0,3}$ are obtained by 5--10\% correction of the former
\cite{gavai}. It has been shown recently that one can extract continuum
results from the lattice spacing we employ \cite{myold}.  Nevertheless,
in future studies we will analyse the effects of relaxing these two
approximations.

Successive configurations used in our computations are separated by
1000 sweeps of a Cabbibo-Marinari heat bath algorithm, so that the
gauge fields are completely decorrelated by any measure one may choose
to use. At $\beta=5.8941$, corresponding to $T=1.5T_c$ we have
generated configurations on $4\times8^3$, $4\times12^3$ and
$4\times16^3$ lattices. At other couplings, corresponding to
$T=0.75T_c$, $1.1T_c$, $1.25T_c$, $2T_c$ and $3T_c$, we have used only the
$4\times12^3$ lattice. We have used quark masses $m/T_c=3$, 1.5, 1,
0.75, 0.30 and 0.03. Using estimates of $T_c$=275--290 for the quenched
theory \cite{myold}, we see that the strange quark mass lies between
$0.3T_c$ and $0.5T_c$.

In this letter we use staggered Fermions exclusively. Since these are
defined for four flavours, we have to normalise the susceptibilities
appropriately \cite{flav}. We have
\beqa
\nonumber
  \chi_0 &=&\frac12({\cal O}_1 + \frac12{\cal O}_2),\\
  \chi_3 &=& \frac12{\cal O}_1,\\
\nonumber
  \chi_s &=& \frac14({\cal O}_1 + \frac14{\cal O}_2),
\label{defs}\eeqa
where the two operators involved are
\beqa
\nonumber
    {\cal O}_1 &=& \frac TV \left\langle{\rm Tr}\,\left(M''M^{-1} 
               - M'M^{-1}M'M^{-1} \right)\right\rangle,\\
    {\cal O}_2 &=& \frac TV 
               \left\langle\left({\rm Tr}\,M'M^{-1}\right)^2 \right\rangle.
\label{ferm}\eeqa
The traces here are sums over lattice points and colour indices, and
angular brackets are averages over gauge field configurations. Primes
denote derivatives of the Dirac matrix with respect to appropriate
chemical potentials. The quark mass to be used in the Dirac operator
for evaluating $\chi_s$ is, of course, different from that for $\chi_{0,3}$.

\begin{figure}[tbh]\begin{center}
   \leavevmode
   \psfig{figure=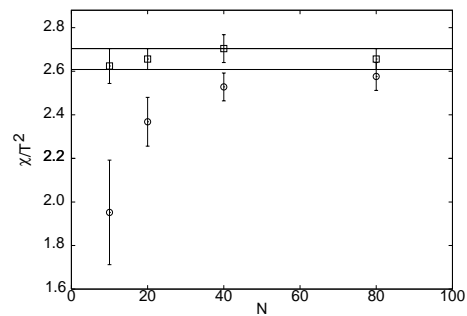,height=4.2cm,width=6cm}
   \end{center}
   \caption{Variation of $\chi_0$ (circles) and $\chi_3$ (squares) with
       the number of vectors, $N$, used per configuration. The averages
       are taken over 20 thermal configurations at $T=1.5T_c$ on a
       $4\times8^3$ lattice for $m/T_c=1$.}
\label{fg.nvec}\end{figure}

The traces are evaluated by the usual stochastic technique,
\beq
   {\rm Tr}\,A = \frac1N\sum_{i=i}^N R_i^\dag AR_i,
\label{stoch}\eeq
where $R_i$ are a set of $N$ uncorrelated vectors with components drawn
independently from a Gaussian ensemble with unit variance. Each vector
has three colour components at each site of the lattice. We improve on
the definitions in eq.\ (\ref{ferm}) by using half lattice versions of
the Dirac operator for staggered Fermions. A detailed discussion of the
stochastic evaluation of the squared trace in eq.\ (\ref{ferm}) can be
found in \cite{gott}.  A systematic study of the optimum value of $N$
is shown in Figure \ref{fg.nvec}. We find that $N\approx80$ is needed
in order to see that $\chi_0=\chi_3$ within statistical errors \cite{ozi}.
In all our subsequent work we have used $N=80$ \cite{milc}. Such a
large value of $N$ also seems to decrease the variance in the average
over gauge configurations.

\begin{figure}[htb]\begin{center}
   \leavevmode
   \psfig{figure=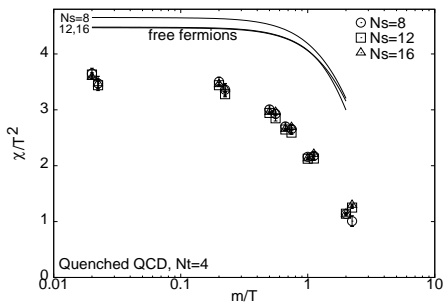,height=4.2cm,width=6cm}
   \end{center}
   \caption{Quark number susceptibilities at $T=1.5T_c$. Values of $\chi_0$
       are displaced slightly to the right for visibility.}
\label{fg.vol}\end{figure}

In Figure \ref{fg.vol} we exhibit the spatial volume dependence of our
results at $T=1.5T_c$. Note that the volume dependence is smaller than
the statistical errors on $\chi$ at this temperature, and
also much smaller than what could be expected for an ideal gas of
Fermions. In view of this, we have chosen to perform the remaining
computations with lattices of size $4\times12^3$. This volume allows us
to measure thermodynamic quantities up to $3T_c$. At larger $T$, spatial
deconfinement sets in and distorts the results unless larger lattices are
used \cite{saumen}. Measurements of finite volume effects on the QCD
equation of state indicates that this lattice size can also be used
down to $1.1T_c$, below which the first order phase transition of the
quenched theory causes strong finite volume effects. As a result, the
quenched theory approximation to full QCD is expected to fail close to
$T_c$.  Another finite size effect appears in Fermion computations in
the quenched theory. As the quark mass decreases at fixed temperature,
the scalar/pseudo-scalar screening length increases. If this length
becomes comparable to the spatial dimensions of the lattice, then
finite volume effects cannot be neglected. Among all our computations,
this happened only for $m/T_c=0.03$ at $T=1.1T_c$.

\begin{figure}[htb]\begin{center}
   \leavevmode
   \psfig{figure=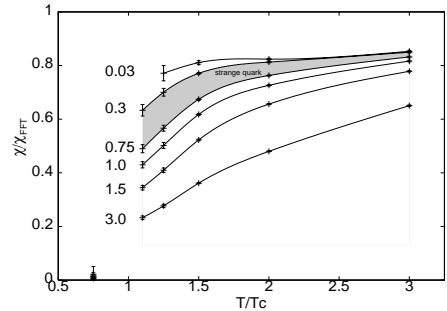,height=4.2cm,width=6cm}
   \end{center}
   \caption{The ratio of $\chi_3$ to its value in a free fermion theory
       at the same lattice volume and quark mass, shown as a function of
       temperature for the values of $m_q/T_c$ indicated. The lines are
       cubic spline fits to the data. The shaded area is the range of
       masses relevant for $\chi_s$.}
\label{fg.tdep}\end{figure}

In Figure \ref{fg.tdep} we collect together all our data. At all these
points, $\chi_0=\chi_3=2\chi_s$ within statistical errors
\cite{note2}.  For $T<T_c$ all the susceptibilities are consistent with
zero. Notice that the results lie significantly below the
expectation from FFT even at temperatures as high as $3T_c$. It is
interesting to note that the departures from an ideal gas become stronger with
increasing quark mass.  We discuss later our checks that this is not a
lattice artifact. The jump in $\chi_s$ across $T_c$, its rapid increase with temperature, and
its becoming comparable with $\chi_{0,3}$ for $T\ge2T_c$, all have
observable consequences in the pattern of relative strangeness
enhancement, including strange baryon enhancement, in going from CERN
to RHIC energies \cite{cleymans}.

It is interesting to speculate upon the reasons for departure from the weak
coupling theory. In view of the long-standing observation that
screening masses in the scalar/pseudo-scalar channel (so-called pion
screening masses) also depart strongly from the weak coupling theory, we would
like to check whether these observations are related. Below $T_c$, at
vanishing chemical potential, pair production of quarks can take place
only by pair production of mesons. The lightest meson, the pion, will
be most effective at producing quark pairs. Hence the pion
susceptibility \cite{sgupta}
\beq
   \chi_\pi = G_\pi(k_0=0,{\bf k}=0) = \frac1{N_z}\sum_z C_\pi(z)
\label{pisus}\eeq
(here $G_\pi$ is the pion propagator in momentum space and $C_\pi$ is
the zero-momentum screening correlator) should determine $\chi_{0,3}$.
For $T>T_c$, this logic does not necessarily follow, but in view of
the strong correlation in the pion sector, it is interesting to test
such a hypothesis.

All the susceptibilities are functions of $m$ and $T$. Since QCD with
two light dynamical flavours is expected to have a second order chiral
phase transition for $m=0$ and $T=T_c$,  near the critical point we
should expect these functions to scale as powers of $T-T_c$ or $m$.
These powers are determined by the ratio of various critical exponents.
The scaled quantities would then vary as universal functions of each
other.  At the critical point, there is a zero mass correlation in the
scalar/pseudoscalar channel which is solely responsible for the
susceptibilities, since the correlations in the other channels are
massive. The universal scaling is then a signal for a restricted form
of dynamical confinement \cite{detar} near the critical point.

\begin{figure}[hbt]\begin{center}
   \leavevmode
   \psfig{figure=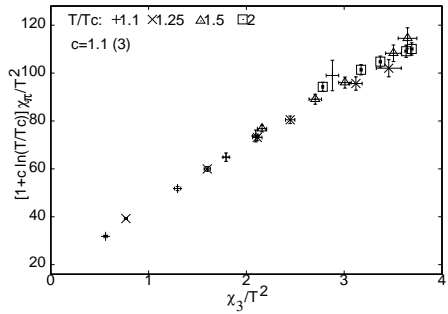,height=4.2cm,width=6cm}
   \end{center}
   \caption{Scaled values of $\chi_\pi$
      %(upper panel) and $\langle\overline\psi\psi\rangle$ (lower panel)
      vary with $\chi_3$
       independently of the values of $T$ and $m/T$ at which they are
       measured, provided $T\le2T_c$ and $m/T_c\le1.5$. The meanings of
       the symbols are the same in the two panels. $\chi_3$ increases
       with decreasing $m/T_c$ at fixed $T$.}
\label{fg.scal}\end{figure}

In quenched QCD, where there is no critical point, we can instead make
an expansion such as
\beq
   \chi_i = a^0_i(T) + a^1_i(T) m + a^2_i(T) m\log m + {\cal O}(m^2),
\label{expn}\eeq
where $i=$ 0, 3, $s$ or $\pi$ \cite{note3}. Eliminating $m$ between
$\chi_\pi$ and $\chi_3$, we can always expand one in terms of the
other, albeit with temperature dependent coefficients.  The surprise,
shown in Figure \ref{fg.scal}, is that the curves for different $T$ can
be scaled to lie on top of each other. The scaling function which
achieves this goes to 1 as $T\to T_c$. Thus, for $T\le2T_c$ and
$m/T\ge1.5$, the co-variance of $\chi_\pi$ and $\chi_3$ at $T_c$
determines that at higher temperatures. Since the scalar/pseudoscalar
screening mass is smallest at finite temperature, and very much smaller
than other screening masses, therefore the observed scaling may be
interpreted as a demonstration of dynamical confinement in much the
same way as at a second order phase transition.

Why is this not a trivial observation? After all, as $m\to0$,
$\chi_\pi/\chi_3$ is a $T$-dependent number, which defines the scaling
function. The point is that the scaling is observed not only for $m=0$,
but for a whole range of masses. This defines an universal curve at
$T_c$, not just a single value. As a result, this scaling also gives us
a partial ability to test for lattice artifacts.  The very fact that
scaling is seen for all $m/T_c\ge1.5$, when results for the lightest
mass are consistent with previous computations in two flavour dynamical
QCD, implies that there are neither strong quenching artifacts, nor
other strong lattice artifacts in our measurements.

In summary, we have presented new and precise results on quark
number susceptibilities over a wide range of temperatures and quark
masses in the high temperature phase of QCD. The susceptibilities
differ significantly from the ideal gas expectations (Figure
\ref{fg.tdep}).  These deviations increase with mass and decrease at
higher $T$. As a result, we expect the relative strangeness enhancement
seen in heavy-ion collisions to increase with temperature as
quantitatively determined here. The linear relation between $\chi_\pi$
and $\chi_3$ (Figure \ref{fg.scal}) is the clearest evidence to date
for the hypothesis of ``dynamical confinement'' in the high
temperature phase of the plasma \cite{detar}. However, it also shows
that such a picture becomes less effective with increasing
temperature.  It is perhaps not a coincidence that the temperature at
which the quenched plasma becomes free of this phenomenon is also the
temperature at which dimensional reduction becomes quantitatively
correct \cite{dimred}. For $T\ge2T_c$, the light quark susceptibility
is known accurately enough to test dimensional reduction and various
other ideas which have emerged in trying to explain lattice results on
the QCD equation of state.

RVG wishes to gratefully acknowledge the kind hospitality of the
Physics Department of Brookhaven National Laboratory where this work
was initiated. SG's work has been sponsored by contract no.\
DE-AC02-98CH10886 of the U.S.\ Department of Energy.

\end{multicols}

\begin{thebibliography}{99}
\bibitem{flav}
  In a quenched computation the number of flavours is defined by
  the valence quarks. Valence staggered Fermions give 4 flavours
  by default, which is accounted for by raising the $det M$ in  
  eq.(\ref{part}) to the power $N_f/4$ for obtaining $N_f$ flavours.
  We have $N_f=2$ of light quarks and one of strange.
\bibitem{ozi}
  It is interesting to note that ${\cal O}_2$, which gives the difference
  between $\chi_3$ and $\chi_0$, is zero in free field theory. It is also
  subleading at large $N_c$.
\bibitem{note2}
  At comparable quark masses and temperatures, there is better
  agreement between the quenched and 2-flavour dynamical QCD results of
  \cite{gott} than seen in the earlier quenched computation \cite{gavai}.
  We believe that the source of this difference lies in spatial volume
  effects close to $T_c$ which affect the quenched and dynamical simulations
  differently.
\bibitem{note3}
  Such an expansion is only valid for $\chi_\pi$ in the high temperature
  phase, as follows from the results of \cite{sgupta}. In the low
  temperature phase, a similar expansion can only be made for $1/\chi_\pi$
  with $a^0_\pi=0$. The importance of the chiral logs is clearest in the
  expansion for the chiral condensate.
\end{thebibliography}
\end{document}